\documentclass[english]{article}
\usepackage[T1]{fontenc}
\usepackage[latin9]{inputenc}
\usepackage{amsmath}
\usepackage{amssymb}
\usepackage{graphicx}

\makeatletter

\providecommand{\tabularnewline}{\\}

\newcommand{\lyxaddress}[1]{
\par {\raggedright #1
\vspace{1.4em}
\noindent\par}
}

\usepackage{babel}

\newcommand{\bra}[1]{\langle#1 |}
\newcommand{\ket}[1]{|#1 \rangle}
\newcommand{\braket}[2]{\left \langle #1 \mid #2 \right \rangle}
\newcommand{\ketbra}[2]{\vert #1 \rangle \! \langle #2 \vert}

\newcommand{\sandwich}[3]{\left \langle #1 \mid #2 \mid #3 \right\rangle}

\newcommand{\Tr}{\mathrm{Tr}}

\usepackage{babel}

\makeatother

\usepackage{babel}
\begin{document}

\title{Propagation of Disturbances in Degenerate Quantum Systems}

\author{Nicholas Chancellor$^{1}$ and Stephan Haas$^{1}$}

\maketitle

\lyxaddress{1 Department of Physics and Astronomy and Center for Quantum Information
Science \& Technology, University of Southern California, Los Angeles,
California 90089-0484, USA}
\begin{abstract}
Disturbances in gapless quantum many-body models are known to travel
an unlimited distance throughout the system. Here, we explore this
phenomenon in finite clusters with degenerate ground states. The specific
model studied here is the one-dimensional J1-J2 Heisenberg Hamiltonian
at and close to the Majumdar-Ghosh point. Both open and periodic boundary
conditions are considered. Quenches are performed using a local magnetic
field. The degenerate Majumdar-Ghosh ground state allows disturbances
which carry quantum entanglement to propagate throughout the system,
and thus dephase the entire system within the degenerate subspace.
These disturbances can also carry polarization, but not energy, as
all energy is stored locally. The local evolution of the part of the
system where energy is stored drives the rest of the system through
long-range entanglement. We also examine approximations for the ground
state of this Hamiltonian in the strong field limit, and study how
couplings away from the Majumdar-Ghosh point affect the propagation
of disturbances. We find that even in the case of approximate degeneracy,
a disturbance can be propagated throughout a finite system. 
\end{abstract}

\section{Introduction}

This paper uses quantum information measures, such as entanglement,
and trace distance to study quantum many body systems. Unlike physical
observables, such quantities usually cannot be directly measured \cite{Cardy 2010},
but can give an important insite into the properties of the system.
Abstract concepts such as quantum entanglement have been important
for almost as long as quantum mechanics has existed \cite{Einstein1935}.
The power of these information theoretical quantities is that they
represent general ideas that can be applied to any system which can
be considered quantum. By studying such abstract quantities one can
more easily generalize a result for a specific system to more universal
behavior. Examples of successful application of quantum information
measures to the study of quantum many body systems are many, a few
examples are \cite{Jacobson2010,Diez2010,Venuti2010(1),Venuti2010(2),Ren2010,Hsu2009}.
The specific uses of these quantities can be diverse, for example
in \cite{Diez2010} the authors use the concept of trace distance
from an averaged density matrix to define a type of quantum equilibration
which would be analogous to equilibration in classical thermodynamics.
Similar questions are examined, but with different methods, in \cite{Venuti2010(1),Venuti2010(2)},
where the concept of equilibration is used to detect criticality in
a system. In \cite{Jacobson2010} a quantity related to fidelity is
used to detect quantum chaos. This paper will make broad use of such
quantum informational quantities, but will deal with relatively few
direct observables. This is because our intention is to provide a
study which can be easily related to other quantum systems, and to
quantum many body theory in general.

The central result of this paper involves a type of local quench which
can propagate disturbances an unlimited distance in a J1-J2 Heisenberg
spin chain. The unitary dynamics of spin chains which can be studied
through quenches can be realized experimentally with trapped cold
atoms \cite{Duan2003,Porras2005}. Certain superconducting qubit arrays
can also provide promising physical realizations of spin chain Hamiltonians
\cite{Makhlin2001,Levitiov2001}. Quenches are also important from
a theoretical perspective. For example, quantum equilibration can
be induced and studied in spin chains using various quenches \cite{Diez2010,Venuti2010(1),Venuti2010(2),Rossini2010}.
Certain local quenches have also been proposed as a way to physically
measure entanglement entropy \cite{Cardy 2010}. Furthermore local
magnetic field quenches similar to those studied in this paper have
been used to study entanglement specifically in Heisenberg spin chains\cite{Ren2010},
as well as other quantum systems \cite{Hsu2009}. A generalization
of the specific system which is studied in this paper has also been
proposed as being possibly useful in quantum computation \cite{Heule2010}.

The frustrated spin-1/2 anti-ferromagnetic Heisenberg chain has one
of the most prototypical matrix product ground states, featuring a
two-fold degeneracy at the so-called Majumdar-Ghosh point \cite{Majumdar1970},
when the nearest-neighbor and next-nearest-neighbor exchange integrals
are the same. The Hamiltonian of this system is given by

\begin{equation}
H_{MG}=\sum_{j=1}^{N}\left(\vec{S}_{j}\cdot\vec{S}_{j+1}+\frac{1}{2}\vec{S}_{j}\cdot\vec{S}_{j+2}\right),\label{eq:hamiltonian}
\end{equation}
 where the sum extends over $N$ lattice sites, and the two terms
represent anti-ferromagnetic nearest-neighbor and next-nearest neighbor
Heisenberg interactions respectively. The ground state of this model
is exactly known \cite{Majumdar1970}, 
\begin{equation}
\ket{\psi_{1,MG}}=\bigotimes_{l=1}^{\frac{N}{2}}\frac{(\ket{\uparrow_{2l-1}\downarrow_{2l}}-\ket{\downarrow_{2l-1}\uparrow_{2l}})}{\sqrt{2}},\label{eq:MDG gs}
\end{equation}
 i.e. the product of nearest-neighbor spin singlets, assuming an even
number of lattice sites. For the case of open boundary conditions,
this state is unique, whereas for periodic boundary conditions it
is two-fold degenerate, as the underlying lattice can be decorated
by the singlet product state in another unique way, 
\begin{eqnarray*}
\ket{\psi_{2,MG}} & = & \bigotimes_{l=1}^{\frac{N}{2}}\frac{1}{\sqrt{2}}(\ket{\uparrow_{mod_{N}2l}\downarrow_{mod_{N}(2l+1)}}-\ket{\downarrow_{mod_{\frac{N}{2}}2l}\uparrow_{mod_{\frac{N}{2}}(2l+1)}}).
\end{eqnarray*}
 The resulting ground state for the periodic system is a superposition,
\begin{equation}
\ket{\psi_{PB,MG}}=a\ket{\psi_{1,MG}}+b\ket{\psi_{2,MG}},\label{eq:MDG gs,pb}
\end{equation}
 where the two terms are not automatically orthogonal. \cite{end1}
Hence, changing the boundary conditions of the Hamiltonian from open
to periodic one goes from a unique to a two-fold degenerate ground
state, thus allowing us to study the effects of a ground state degeneracy.

Local disturbances of this ground state can be introduced by applying
a local magnetic field $h$ to a subset of $N'$ adjacent spins, 
\begin{equation}
H(h,N')=H_{MG}-h\sum_{j=1}^{N'}S_{j}^{z},\label{eq:field}
\end{equation}
 where without loss of generality we consider the direction of the
applied field to be along the z-direction. 

One can take advantage of the fact that spin polarization is conserved
in this system, allowing one to reduce the complexity of the problem
by dividing the Hamiltonian into independent spin sectors, which may
each be diagonalized independently. These sectors correspond to the
total polarization of the system in the z direction, and may be diagonalized
independently. The polarization sector which contains the global ground
state of the system changes with field strength, therefore figures
\ref{fig:sd_&_L_vs_h}, \ref{fig:sd_&_L_vs_h},\ref{fig:open SDvsEM},\ref{fig:periodic SDvEM},
\ref{fig:singlet shift}, and \ref{fig:initial dist to av} all show
curves for three different polarization sectors. Each sector is labeled
with the total z polarization of the entire spin chain in that sector,
which is conserved under the action of all Hamiltonians considered
in this paper. For example in the basis where $S_{j}^{z}$ is diagonal,
all of the basis states in the L=0 sector will have the same number
of spins pointing in +z as -z, in the L=-1 state, 2 more spins will
be facing in -z than +z, etc.

\begin{figure}[h]
 \includegraphics[scale=0.4]{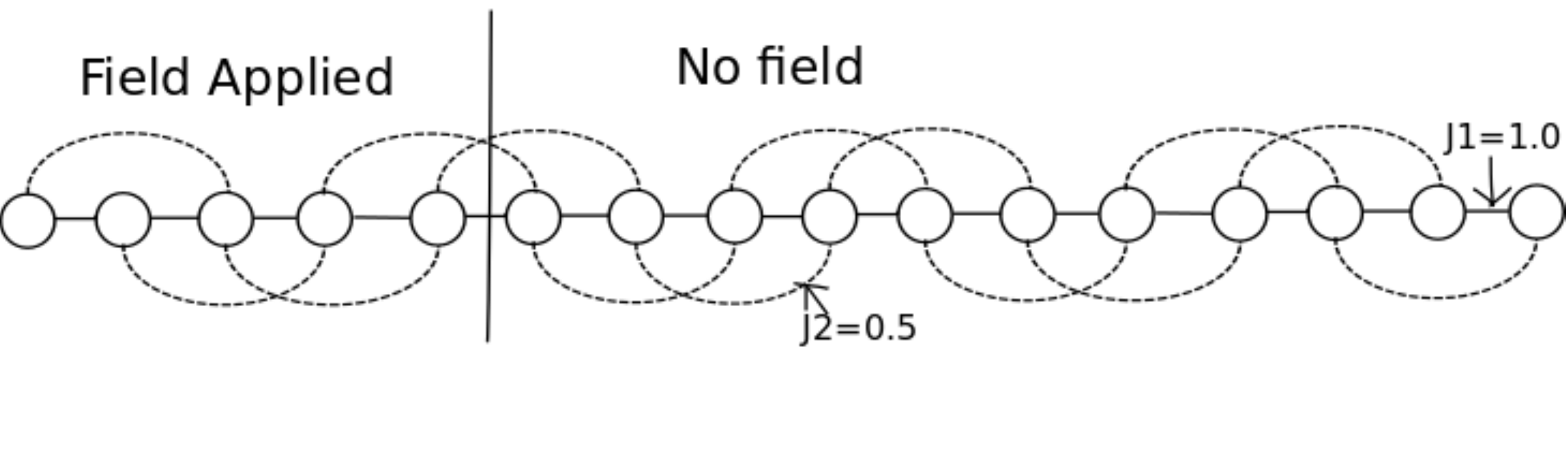}\label{fig:quench} \caption{Example of a local field applied to the Majumdar-Ghosh Hamiltonian.}
\end{figure}

In this study, we identify several effects induced by the application
of a local magnetic field, as depicted in Fig. \ref{fig:quench}.
Here we briefly summarize our findings.

Firstly, for sufficiently small field amplitudes polarization induced
by the local magnetic field is stored in the vicinity of the region
to which the field is applied, instead of spreading throughout the
entire system. Only beyond a certain threshold field, i.e. once some
of the polarization in this boundary region has saturated, can it
spread throughout the entire system. We argue that this is to be expected
because at the Majumdar-Ghosh point the energy spectrum of the J1-J2
Heisenberg Hamiltonian is gapped. Provided that the energy gained
from the locally applied magnetic field is small compared to the coupling
energy of the spins, any state which keeps the majority of spins in
a matrix product configuration similar to the zero field ground state
will have a lower energy. For an even number of spins in the non-field
region, the system can only accomplish this if the total polarization
of a given subsystem far from the field region is zero. The spins
in the field region align in the direction of the applied field, thus
in turn leading to an excess opposite polarization of the spins not
directly subjected to the field. This induced polarization is typically
localized near the edge of the field region. We will show, however,
that this effect does not occur if the two degenerate ground states
lie in different polarization sectors, because in this case the polarization
can spread through the degenerate subspace at no energy penalty.

\begin{figure}
\includegraphics[scale=0.4]{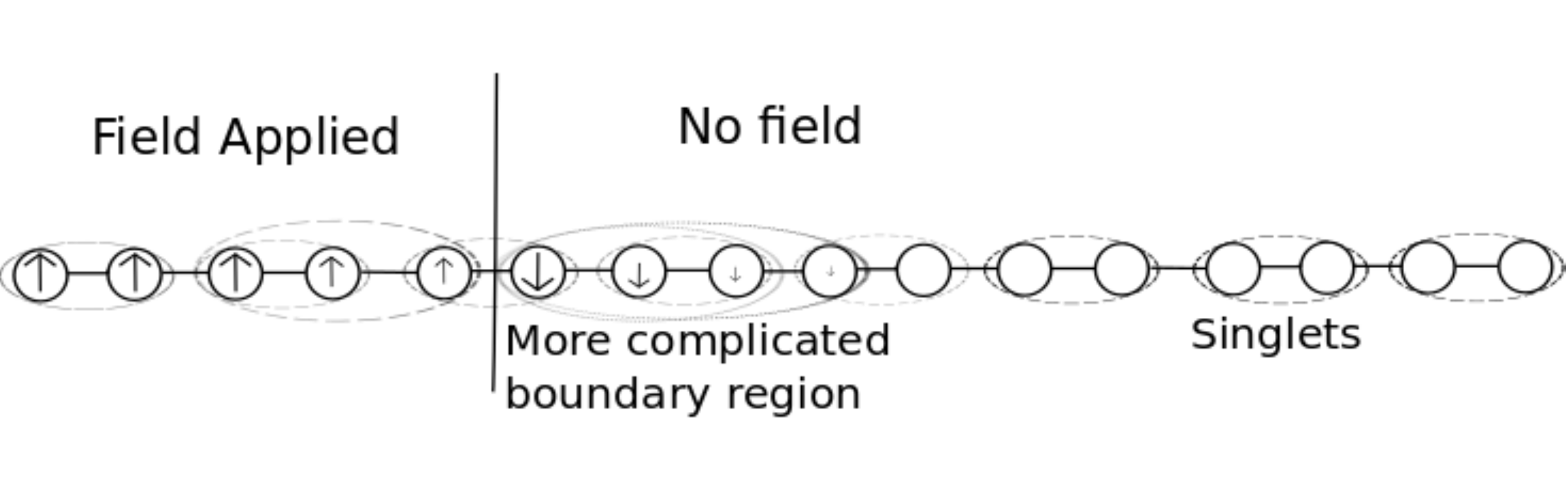}

\caption{Sketch of a typical state of the spins when exposed to a field. For
all field strengths studied here, the ground state of at least one
total spin sector behaves like this, and one of these states always
is the global ground state of the system. Ovals represent entanglement,
arrows indicate spin polarization.}

\label{fig:boundry cartoon} 
\end{figure}

We will also show that, for a sufficiently small fraction of the spins
subjected to the field, there exists at least one state in one of
the polarization sectors which looks locally like the zero field (MPS)
ground state far from the field (Fig. \ref{fig:boundry cartoon}).
For the systems studied in this paper one of these states is always
the ground state. \cite{end2}

For the case of periodic boundary conditions, any state which lies
locally in the degenerate subspace far from the local magnetic field
region will have the minimum local contribution to the energy. This
means that even for a system with many more spins outside of the field
than within it, a disturbance can easily propagate throughout the
entire zero field region.

This paper is organized as follows. In the following section 2, we
introduce the observables on which we focus to understand the effects
of a local applied magnetic field on this many-body system. The cases
of open and periodic boundary conditions need to be treated separately.
In section 3, we then discuss the physics of open chains, and in section
4 the phenomena observed in periodic systems. In section 5, we consider
how these results are affected when one departs from the Majumdar-Ghosh
point in the underlying Hamiltonian. This is followed by conclusions
in section 6.

\section{Physical observables}

\subsection{Open boundary conditions}

For open boundary conditions the field is applied to N' spins on one
end of the chain. Unless otherwise stated, we consider finite chains
with a total number of spins, N, performing full numerical diagonalizations
of the frustrated Majumdar-Ghosh Heisenberg Hamiltonian.\cite{end3}
Several observables are studied. The first is the total polarization
outside of the region subjected to the applied field. While the total
spin polarization of the chain is conserved, local polarization is
not. This quantity is defined as 
\begin{equation}
L_{\neg N'}=\sandwich{\psi}{{\textstyle \sum}_{j=N'+1}^{N}S_{j}^{z}}{\psi}\label{eq:L not N'}
\end{equation}

Furthermore, we study the trace distance from a singlet state of the
two spins at the end of the chain opposite to the region of the applied
magnetic field, i.e. the spins located at sites $N-1$ and $N$. This
observable is defined as 
\begin{eqnarray}
d_{s}=\frac{1}{2}\Vert\rho_{s}-\frac{1}{2}(\ket{\uparrow\downarrow}-\ket{\downarrow\uparrow})(\bra{\uparrow\downarrow}+\bra{\downarrow\uparrow})\Vert_{1},\label{eq:sing dist}
\end{eqnarray}
 where 
\begin{eqnarray}
\rho_{s} & = & \Tr_{\neg s}(\ketbra{\psi}{\psi}),\\
\Vert O\Vert_{1} & \equiv & \Tr\sqrt{O^{\dagger}O}.\label{eq:trace norm}
\end{eqnarray}

Finally, we focus on the polarization of the spins at sites $N-1$
and $N$, defined the same as in Eq. \ref{eq:L not N'}, but with
the sum running from N-1 to N. In this paper the subsystem of the
2 furthest spins will be labeled f. This observable tells about whether
the polarization has been allowed to spread to the furthest 2 spins
from the field.

Two different sizes of field regions are considered, N'=5 and N'=4.
The reason that both are considered separately is that there are significant
even-odd effects.

In this paper, no actual quenches are performed in the system with
open boundary conditions, and all observables are given for the ground
state of a given sector.

\subsection{Periodic boundary conditions}

For periodic boundary conditions, the field is applied to a region
of N' adjacent spins. In this case, we are considering chains with
an even number of sites. While the observables studied in the periodic
case are defined in analogy to those studied in the open case, some
extra care is necessary. In particular, a complication arises for
the trace distance from a singlet for the two spins furthest from
the field region. For periodic boundary conditions, there is no unique
choice of singlet covering for the system. Two different approaches
to this problem are examined. Firstly, one can consider the distance
from the closest of the two singlet coverings for a subsystem, 
\begin{eqnarray}
d_{s,cover}=min(\Vert\rho_{s}-\Tr{}_{\neg s}(\ketbra{\psi_{1,NF}}{\psi_{1,NF}})\Vert_{1},\nonumber \\
\Vert\rho_{s}-\Tr{}_{\neg s}(\ketbra{\psi_{2,NF}}{\psi_{2,NF}})\Vert_{1}).\label{eq:dist sing cover}
\end{eqnarray}
 However, this quantity has a drawback, i.e. all but a zero measure
set of states in the degenerate subspace will have a finite distance
to either of these coverings. An alternative approach is to look at
the distance from the closest point in the subspace to the reduced
density matrix, 
\begin{eqnarray}
d_{s,subspace}=min_{a,b}(\Vert\rho_{s}-(\Vert a\ket{\psi_{1,NF}}+b\ket{\psi_{2,NF}}\Vert_{2})^{-2}\times\nonumber \\
(\Tr[(a\ket{\psi_{1,NF}}+b\ket{\psi_{2,NF}})(a^{\dagger}\bra{\psi_{1,NF}}+b^{\dagger}\bra{\psi_{2,NF}})]\Vert_{1}).\label{eq:dist sing subspace}
\end{eqnarray}
 This equation appears as though it can be further simplified in an
obvious way, but remember that the two wave functions are not orthogonal.
The norm in the denominator is the usual L2 norm for a vector. Also
in this case the minimization is actually simpler than it looks, by
realizing that it can be reduced to: $d_{sing,subspace}=min_{0\leq\alpha\leq1}\Vert\rho_{s}-((1-\alpha)\times\frac{1}{2}(\ket{\uparrow\downarrow}-\ket{\downarrow\uparrow})(\bra{\uparrow\downarrow}+\bra{\downarrow\uparrow})+\alpha\times1_{4})\Vert_{1}$
where $1_{4}$is the 4-dimensional identity operator.

While the observables presented in this section could be considered
as time dependent variables, in this paper they are always studied
for the ground state of a given polarization sector.

\subsection{Small magnetic field quenches}

Because of the degeneracy caused by the periodic boundaries there
is another quantity which is interesting to look at, relating to a
field quench performed by changing the magnetic field instantaneously
and subsequently monitoring the time evolution of the system, especially
in regions far from where the local field is applied. Unitary evolution
gives the time evolution of a system following a quench at time $t=0$,
in terms of energies $E_{n}$, 
\begin{equation}
\rho_{m,n}(t)=c_{m}^{*}c_{n}\exp[-\imath(E_{n}-E_{m})t]\quad,\label{eq:time evolution}
\end{equation}
 where $c_{m}=\braket{m}{\psi}$, where $\ket{\psi}$is the pre-quench
ground state of the system. This leads to a definition of the time
averaged state, 
\begin{equation}
\bar{\rho}_{m,n}=c_{m}^{*}c_{n}\delta(E_{n}-E_{m})\quad.\label{eq:rho bar}
\end{equation}

The field quench is performed by taking $\ket{\psi(t=0)}=\ket{\psi_{0}}$
to be the ground state of a Hamiltonian with a slightly stronger field,
$H_{0}=H-\epsilon\sum_{j=1}^{N'}S_{j}^{z}$. At $t=0$, $\epsilon$
is instantaneously turned off. In our analysis of the time evolution,
we will focus on the trace distance from the time averaged (or dephased)
state of the density matrix of the two spins furthest away from the
field region 
\begin{equation}
d_{av}(t)=\Vert\Tr{}_{\neg s}(\ketbra{\psi(t)}{\psi(t)})-\bar{\rho}_{s}\Vert_{1}.\label{eq:dist av}
\end{equation}

\subsection{Large magnetic field quenches}

We also examine the time evolution due to large local field quenches.
Several statistical distributions are studied to understand the ensuing
equilibration behavior. These quenches are performed in a regime where
quenches are shown to disturb the entire system, even regions far
away from the field region. A global quantity which is studied is
the Loschmidt echo, a measure of the overlap of the time evolved system
with the initial state, 
\begin{equation}
LE(t)=|\sandwich{\psi}{exp(-\imath\, H\, t)}{\psi}|^{2}.\label{eq:Loschmidt}
\end{equation}

Two local linear quantities are examined as well. In the region subjected
to the local external field, the local polarization is studied. This
is simply the expectation value of the magnetization operator with
respect to the local density matrix, 
\begin{equation}
L_{N'}(t)=\Tr(\rho_{N'}(t)M).\label{eq:locMag}
\end{equation}

In the region far from the spins where the local magnetic field is
applied, all of the states are expected to be locally within the degenerate
ground state subspace and therefore have zero magnetization. Therefore,
a more appropriate observable to use is the overlap with a singlet
state, 
\begin{equation}
O_{s}(t)=\Tr(\rho_{s}(t)\Tr{}_{\neg s}(\ketbra{\psi_{1,NF}}{\psi_{1,NF}})).\label{eq:singlet overlap}
\end{equation}

Finally an important non-linear local quantity is studied far from
the local magnetic field, the time evolving distance to the average
state, defined by 
\begin{equation}
d_{s}(t)=\Vert\rho_{s}(t)-\bar{\rho}_{s}\Vert_{1}.\label{eq:time dist av}
\end{equation}
 This quantity is important, as it provides a direct measure of equilibration
locally, and can thus be used to show that the quench not only disturbs
the system far from the field, but also that these disturbances can
cause equilibration.

\subsection{Entanglement maps}

A tool which is used in this paper for visualizing quantum states
is a map of two point entanglement. In these graphics, colors are
used to indicate entanglement strength between single spins using
von Neumann entropy, 
\begin{equation}
S_{VN}(\rho)\equiv\Tr(\rho\log(\rho)),\label{eq:svn}
\end{equation}
 as a measure of entanglement.

These graphics consist of arrays of colored squares where, for off-diagonal
elements, the color corresponds to the entanglement between the two
spins. The diagonal elements correspond to the difference between
the maximum possible entropy on a spin and the actual entropy. This
represents the the amount of information left about a spin after the
rest of the system is measured. These maps are created using 
\begin{eqnarray}
entMap(i,j) & = & (1-\delta_{ij})*((S_{VN}(\rho_{i})+(S_{VN}(\rho_{j})-S_{VN}(\rho_{ij}))+\label{eq:entMap}\\
 &  & \delta_{ij}*(S_{VN}(\frac{1}{2}*1_{2})-S_{VN}(\rho_{i}))\nonumber 
\end{eqnarray}

The color scale with the maximum entanglement normalized to 1 appears
in Fig. \ref{fig:EntMap L1 h5 s4}. It is important to note that while
these figures can give a good general impression of entanglement behavior
of the system, they do not tell the whole story, i.e. they only give
information about two-point entanglement. Just because one of these
figures shows no two point entanglement for a pair of spins, this
does not mean that they are not entangled in a more complicated way.\cite{end4}

Although in principle there is nothing preventing one from obtaining
entanglement maps for time averaged states, in this paper we only
use this technique to study eigenstates.

\section{Local magnetic field applied to Majumdar-Ghosh chains with open boundaries}

\begin{figure}[h]
\includegraphics[scale=0.4]{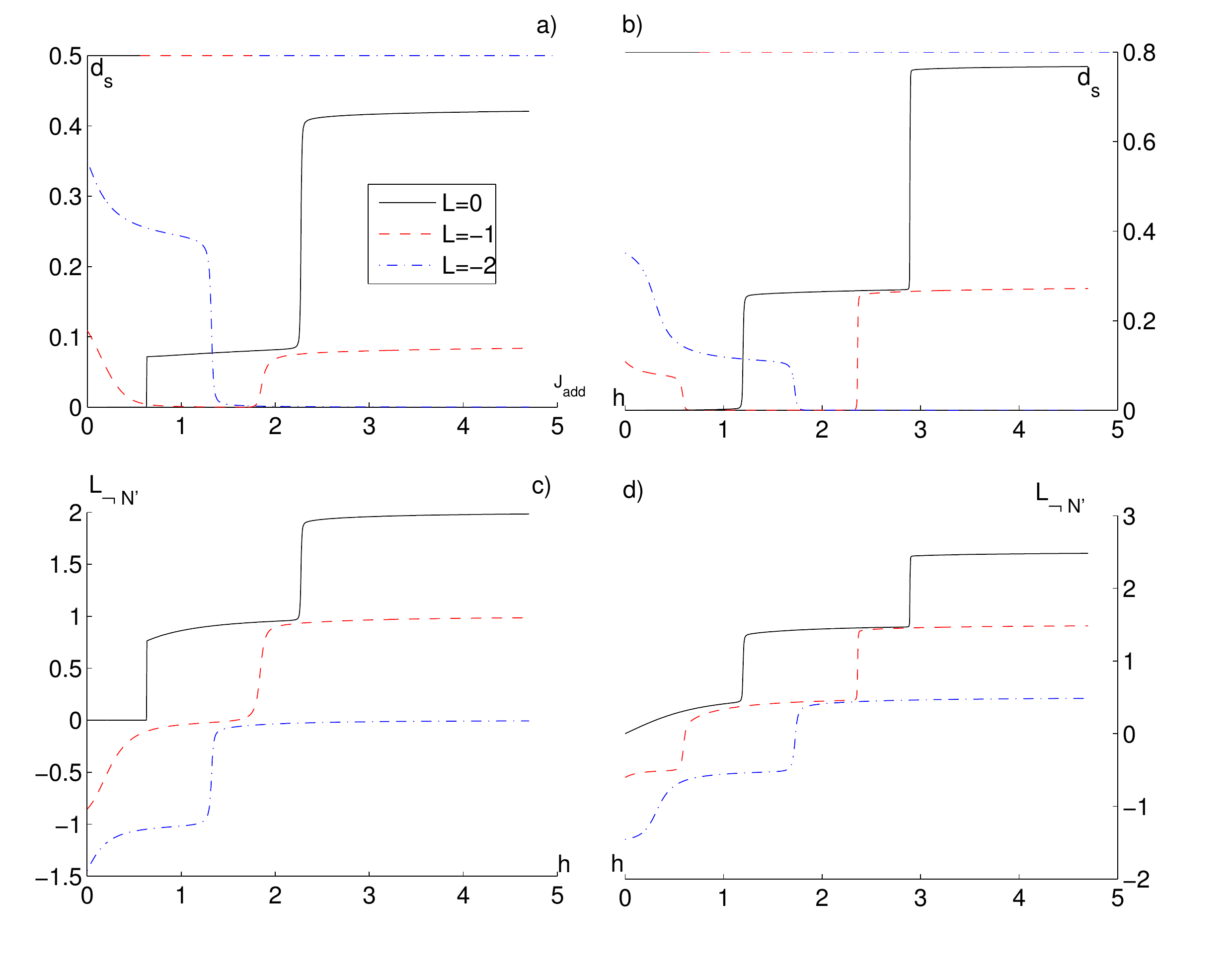} \caption{(color online) (a) and (b): Trace distance from a singlet state of
the two spins at the end of the chain opposite to the region subjected
to the local magnetic field. (c) and (d): Total spin polarization
outside of the region subjected to the local field. (a) and (c) are
for local fields applied to 4 spins, and (b) and (d) are for local
fields applied to 3 spins. On all figures, the solid line is the L=0
sector, dashed lines indicate the L=-1 sector and the dot dashed lines
indicate the L=-2 sector. Note that for sufficiently small local fields,
the global ground state lies in the L=0 sector, whereas for larger
local field strengths it lies in higher polarization sectors. In both
cases the global ground state is locally close to the singlet state
on spins far from the field. These plots are all properties realting
to the ground states of given sectors.}

\label{fig:sd_&_L_vs_h} 
\end{figure}

Subjecting a local region of a Majumdar-Ghosh spin chain to an external
magnetic field forces the exposed spins to align with the field. Because
of polarization conservation, excess polarization opposite to the
direction of the field is generated in the field-free region of the
system. In the sector of zero total spin polarization ($L=0$), and
for sufficiently large magnetic field strengths, this can cause spins
far from the field region to switch to non-trivial polarized configurations,
whereas for smaller applied fields they remain in a spin singlet product
state. In contrast, in polarization sectors with $L\neq0$ excess
polarization is trapped close to the region where the field is applied,
and singlets are pushed far away from this field region. This is demonstrated
graphically in Fig. \ref{fig:sd_&_L_vs_h} where parts (a) and (b)
show the trace distance of two spins far from the locally applied
magnetic field from a singlet for fields on an even and odd number
of spins respectively. Parts (c) and (d) show the polarization stored
in the region with no applied magnetic field versus field strength,
again for fields on an even and odd number of spins respectively.
As Figs. \ref{fig:sd_&_L_vs_h}(a) and (b) show, for local fields
applied to regions with both an even and odd number of spins, there
is always at least one polarization sector for which singlets are
located far away from the field region. Even for relatively small
finite systems, such as the ones studied here, the ground state always
lies in one of these sectors.

It is also interesting to note from Figs. \ref{fig:sd_&_L_vs_h}(c)
and (d) that for a small field in the L=0 sector, the spins in the
field-free area behaves differently, depending on whether the local
field is applied to an odd or to an even number of spins. This can
be explained by the fact that for a field on an odd number of spins,
the boundary between the field and the region with no field cuts through
a singlet, in the original ground state. E.g. the field gradient makes
one component of the singlet more energetically favorable than the
other. By rotating these two spins between the singlet and the classical
$\ket{\downarrow\uparrow}$ state, the ground state can be adjusted
locally. When, however the field boundary is between two singlets,
a critical local field strength must be reached for any polarization
to be transferred from the field region to the field-free region as
Fig. \ref{fig:sd_&_L_vs_h}(c) demonstrates. This is because the matrix
product state of singlets is still an eigenstate of the Hamiltonian
for any field strength in this case, and a level crossing must occur
before the ground state can change. \cite{end5}

\begin{figure}
\includegraphics[scale=0.5]{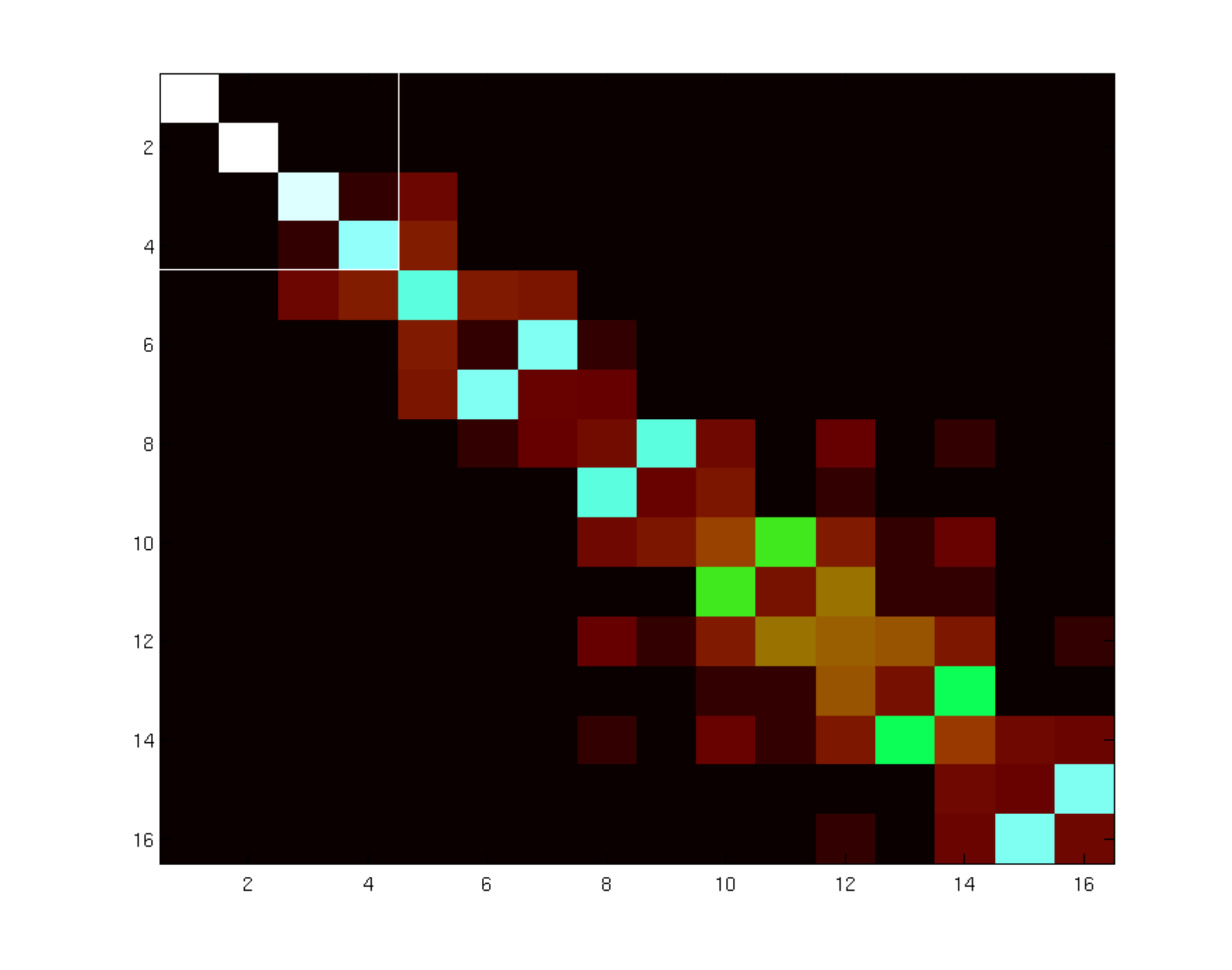}

\includegraphics[bb=-62bp 0bp 505bp 62bp,scale=0.5]{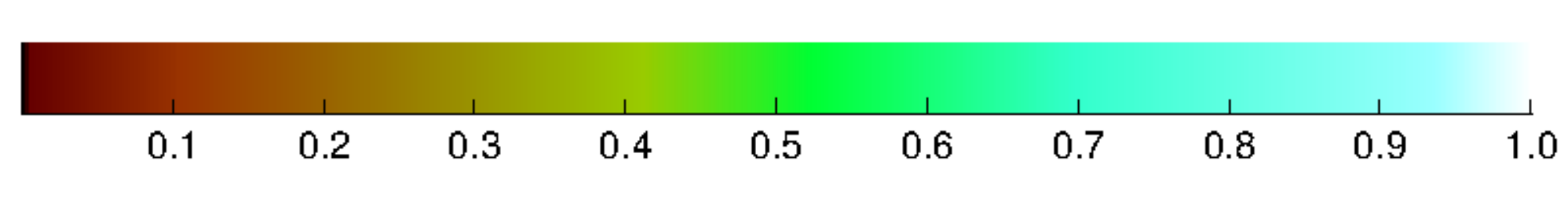}

\caption{(color online) Entanglement map for the L=-1 sector ground state (note
that this is not the global ground state) for a Majumdar-Ghosh chain
of 16 spins with 4 adjacent spins, whose position is indicated by
the white square, subjected to a local magnetic field of strength
h=5J . The color scale is normalized to 1 as shown.}

\label{fig:EntMap L1 h5 s4} 
\end{figure}

Fig. \ref{fig:EntMap L1 h5 s4} shows the entanglement map of a system
in the L=-1 global spin sector, with a magnetic field of h=5J applied
on 4 of 16 spins. Fig. \ref{fig:EntMap L1 h5 s4} suggests that for
a range of field values, the distance from a singlet is caused by
frustration from having an effectively odd number of spins available
in the Majumdar-Gosh Hamiltonian. In this case, however, the frustration
is alleviated by an intermediate transition region between the field
behavior and far from field behavior changing its length (at the cost
of some energy). \cite{end6}

\subsection{Polarization effects}

\begin{figure}
\includegraphics[scale=0.4]{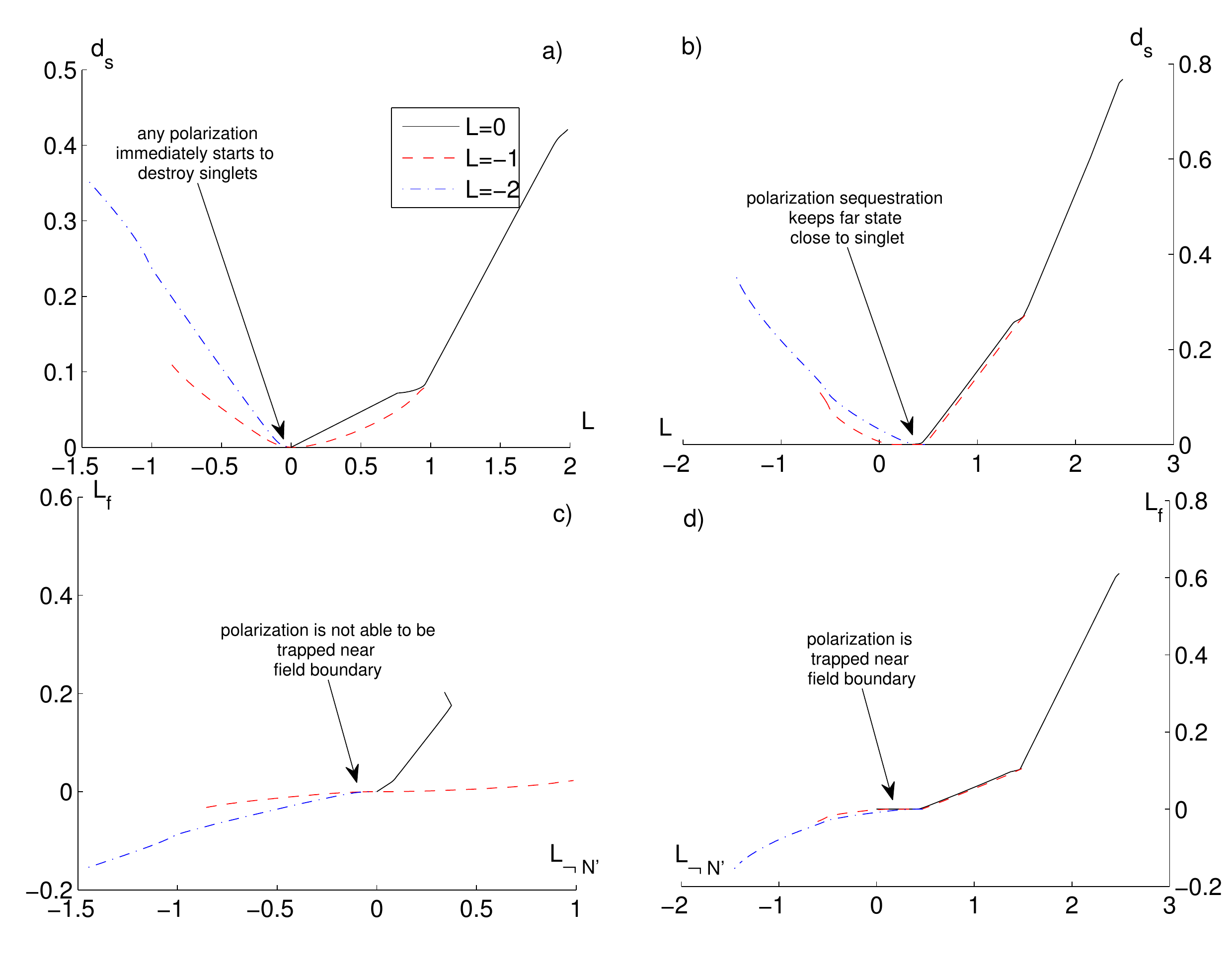}

\caption{(color online) (a) and (b): Trace distance from a singlet state of
the two spins at the end of the chain opposite to the region subjected
to the local magnetic field versus polarization of the entire field
free region. (c) and (d): Spin polarization of the 2 furthest spins
versus polarization of the entire field free region. Local field on
4 of 16 spins with periodic boundary conditions (left column). Local
field on 5 of 16 spins with periodic boundary conditions (right column).
In all parts, the solid line is the L=0 sector, dashed lines indicate
the L=-1 sector and the dot dashed lines indicate the L=-2 sector.
These plots are all properties realting to the ground states of given
sectors.}

\label{fig:open SDvsEM} 
\end{figure}

The way the system distributes polarization depends strongly on even-odd
effects. To study the effects of polarization we examine Fig. \ref{fig:open SDvsEM}
which shows the dependence of trace distance from a singlet for spins
far from the locally applied magnetic field on the polarization in
the non-field region in parts (a) and (b) for fields on an even and
odd number of spins respectively. Parts (c) and (d) show the polarization
of the last 2 spins rather than trace distance from a singlet. From
Figs. \ref{fig:open SDvsEM}(c) and (d) one can tell that if the field
is placed on an even number of spins, any polarization that is in
the non-field region will be immediately spread, even to the furthest
spins. In the case where the field is placed on an odd number of spins,
however, a finite amount of polarization can be sequestered near the
boundary. Figs. \ref{fig:open SDvsEM}(a) and (b) show that this trend
is mirrored in distance from a singlet for far-away spins.

\begin{figure}
\includegraphics[scale=0.4]{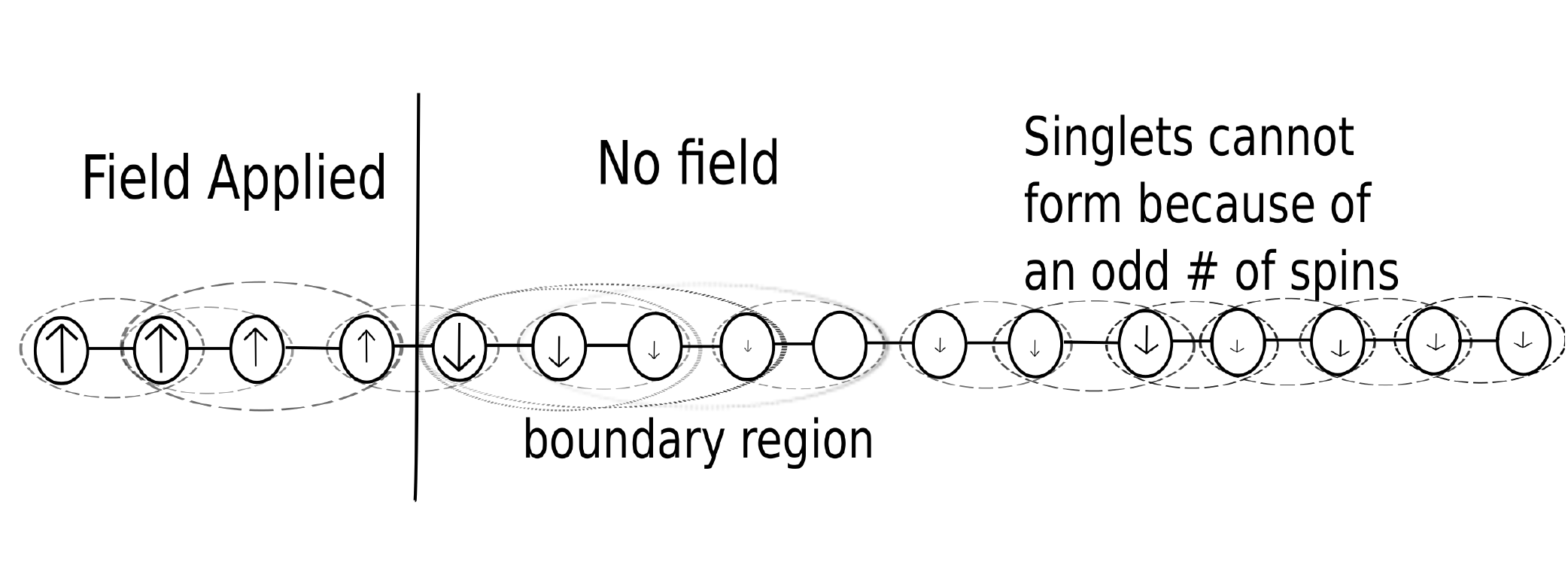}

\caption{Cartoon representation of the effect which prevents polarization sequestration
for a field on an even number of spins.}

\label{fig:frustration cartoon} 
\end{figure}

The differences between the even-spin and odd-spin ground state for
the zero-field spin chain can be used to explain why polarization
sequestration can occur in one case and not the other. Any spin $\frac{1}{2}$
spin chain with an odd number of spins and no applied local field
must have a degenerate ground state because the particle-hole duality.
The degenerate ground states also have different polarization and,
therefore, 2 degenerate ground states with a continuum of polarization
between $L=-\frac{1}{2}$ and $L=\frac{1}{2}$ are possible. This
means that for a chain which is effectively {}``odd'', there is
no energy penalty for being anywhere in this range. This effect allows
polarization to be spread throughout the no-field region without increasing
the energy in that region. Polarization can effectively be moved through
this locally degenerate subspace, therefore polarization sequestration
does not occur. Conversely, for a spin chain which is effectively
{}``even'', the ground state is unique, and polarization will tend
to be localized in the ground state to avoid raising the energy of
all of the no-field spins. As Fig. \ref{fig:EntMap L1 h5 s4} suggests,
for certain field ranges in a given sector, the length of the non-field
region of the chain is effectively {}``odd''. When this happens
polarization can be spread freely throughout the non-field region,
and sequestration does not occur, see Fig. \ref{fig:frustration cartoon}.

\subsection{Field Induced Effects}

\begin{figure}
\includegraphics[scale=0.4]{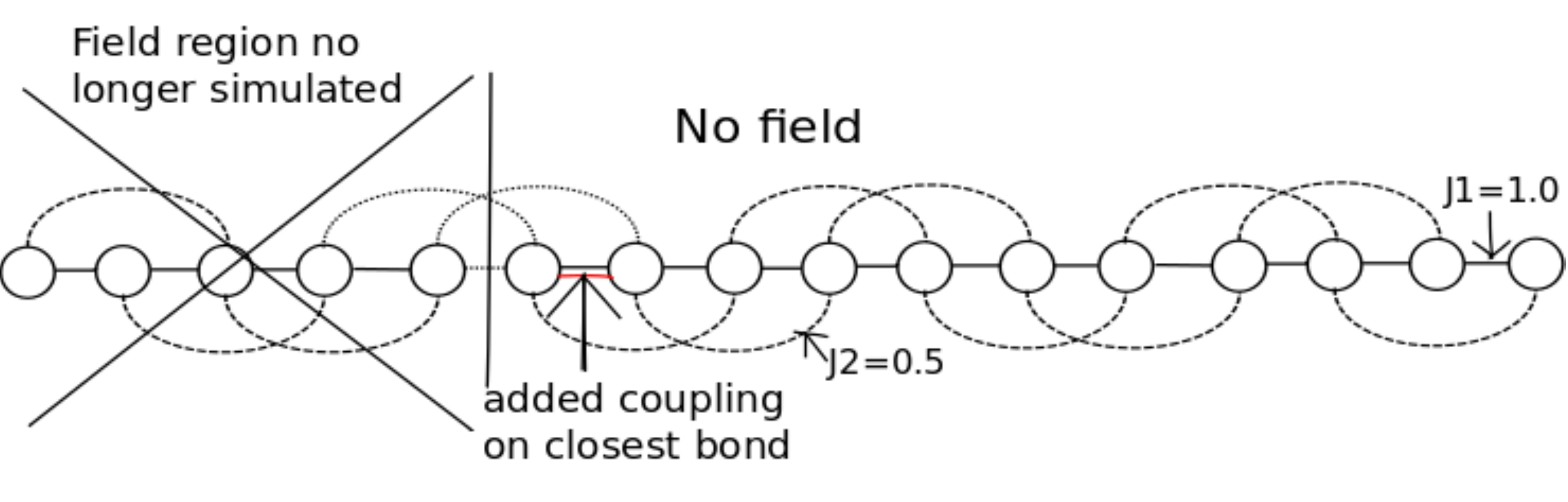}

\caption{(color online) The approximation used to simulate behavior with a
strong field.}

\label{fig:coupling approx cartoon} 
\end{figure}

\begin{figure}
\label{fig:coupling approx gs}\includegraphics[scale=0.5]{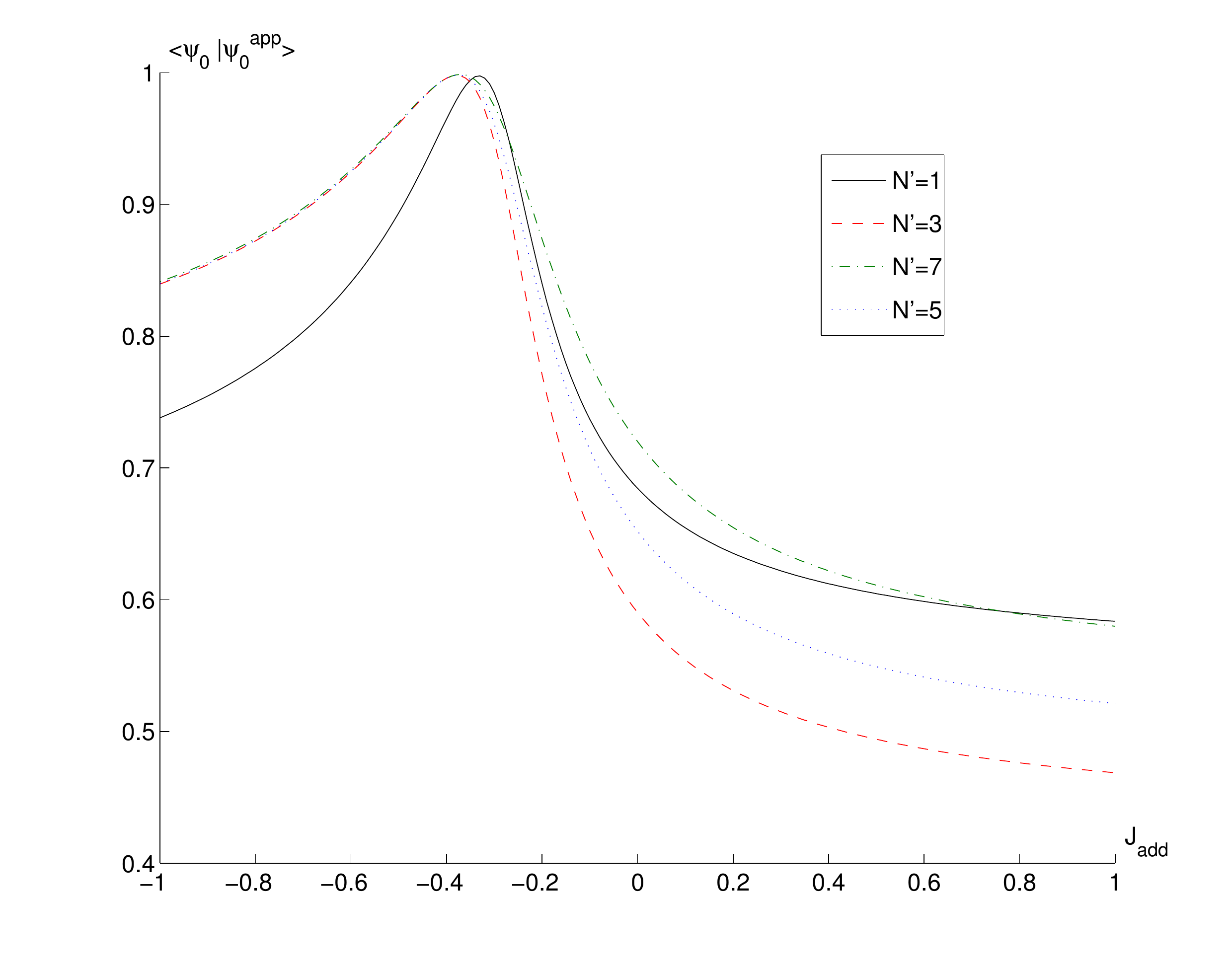}

\caption{(color online) Overlap between actual ground state, and ground state
of a Hamiltonian which is applied only to the non-field spins (tensored
with spins opposing the field in the field region), but with a modified
coupling on the two spins closest to the field. X axis is the additional
coupling added to the Majumdar-Ghosh Hamiltonian. Data was taken with
h=100, N=16 (open boundaries). Different lines are as follows solid-N'=1,
dashed-N'=3, dotted-N'=5, dot dashed-N'=7. Even N' (not shown) are
not accurately represented by this model.}
\end{figure}

For very strong fields, the spins within the field should have no
entanglement with the rest of the system, in low energy states. This
is because the spins subjected to the field will align with the field.
Therefore an effective Hamiltonian which acts only on the spins outside
of the field should be able to describe the system in low energy states.
A simple model for this Hamiltonian would be to alter the coupling
between the two spins closest to the field, with the supposition that
the coupling with the field spins acts to mediate the interaction
between the two spins coupled to them (see Fig. \ref{fig:coupling approx cartoon}).
The overlap between the known ground state, and the ground state calculated
using the approximation shown in Fig. \ref{fig:coupling approx cartoon}
for different added coupling strengths and different spins in the
field region appear in Fig. \ref{fig:coupling approx gs}. Fig. \ref{fig:coupling approx gs}
supports the claim that this approximation works fairly well in the
ground state for a field on an odd number of spins. For numerical
results see table \ref{tab:coupling approx}.

\begin{table}
\begin{tabular}{|c|c|c|}
\hline 
\multicolumn{1}{|c||}{} & \multicolumn{1}{c||||}{$\braket{\psi_{0}}{\psi_{0}^{app}}$} & $J_{add}$\tabularnewline
\hline 
\hline 
N'=1  & 0.9976  & -0.3323\tabularnewline
\hline 
N'=3  & 0.9980  & -0.3786\tabularnewline
\hline 
N'=5  & 0.9983  & -0.3756\tabularnewline
\hline 
N'=7  & 0.9987  & -0.3706\tabularnewline
\hline 
\end{tabular}

\caption{Statistics considering a field of h=100 placed on N' spins, comparing
the approximate to the actual Hamiltonian. The coupling listed here
is the additional coupling added to the 2 closest spins to the field}

\label{tab:coupling approx} 
\end{table}

\section{Local magnetic field applied to Majumdar-Ghosh chains with periodic
boundary conditions}

\begin{figure}
\includegraphics[scale=0.4]{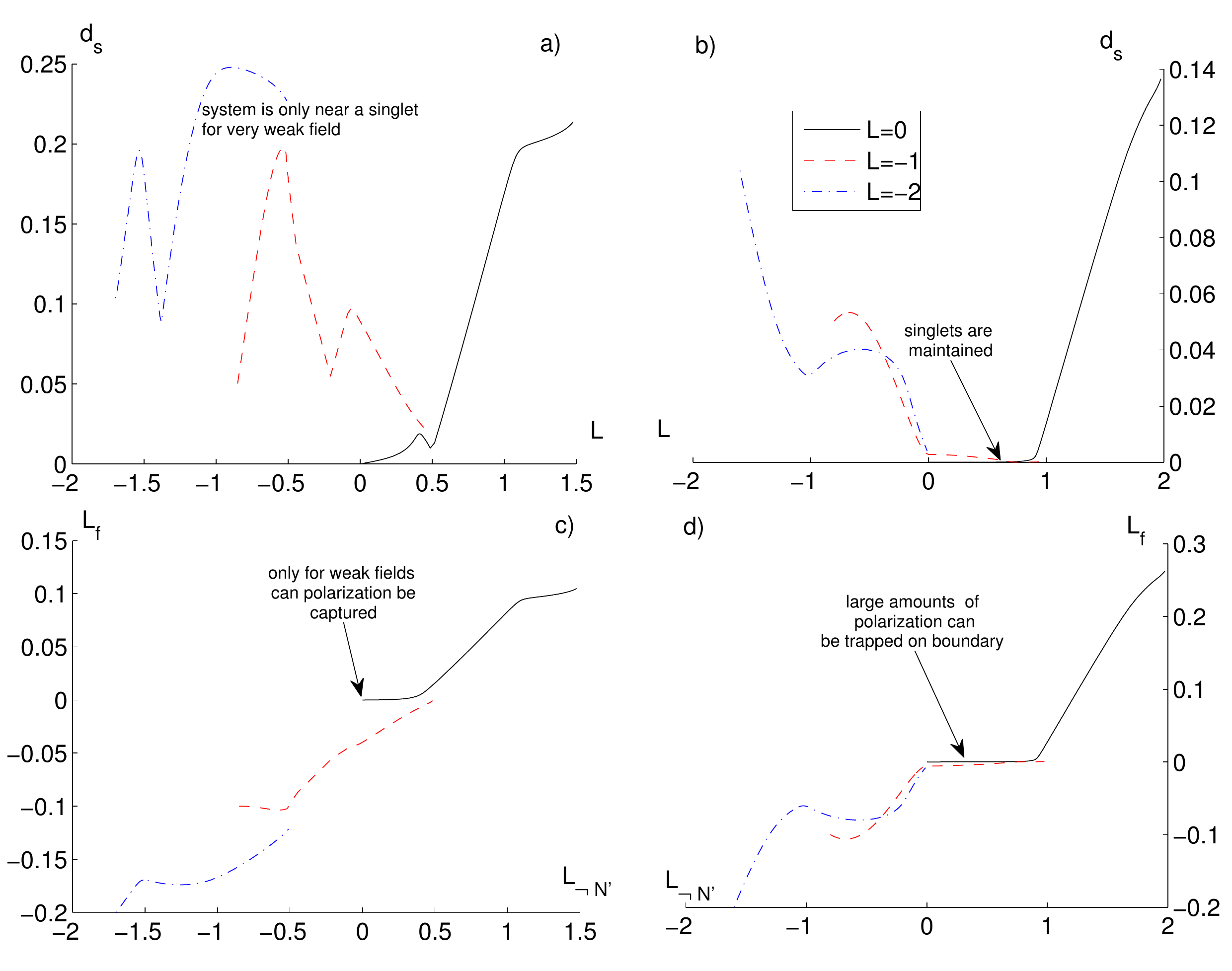}

\caption{(color online) Closest local distance from singlet subspace of 2 furthest
spins from the locally applied magnetic field versus polarization
on all non-field spins (top row). Polarization on 2 furthest spins
versus polarization on all non field spins (bottom row). Field on
3 of 20 spins with periodic boundary conditions (left column). Field
on 4 of 20 spins with periodic boundary conditions (right column).
On all figures, the solid line is the L=0 sector, dashed lines indicate
the L=-1 sector and the dot dashed lines indicate the L=-2 sector,
where I call negative L to be in the direction of the field. These
plots are all properties realting to the ground states of given sectors.}

\label{fig:periodic SDvEM} 
\end{figure}

Unlike open boundary conditions, periodic boundaries present a case
where the unperturbed Hamiltonian has a degenerate ground state. Therefore,
the local Hamiltonian for the spins far away from the region subjected
to the local field will also always have a degenerate ground state.
The complications from this degeneracy add a new series of effects
which are not observed in the open-boundary case. These effects are
illustrated by Fig. \ref{fig:periodic SDvEM} which shows in parts
(a) and (b) the closest distance from the singlet subspace for the
two furthest spins from the region of the locally applied magnetic
field versus polarization on all non-field spins for 3 of 20 and 4
of 20 spins in the field respectively. Parts (c) and (d) show polarization
on the two furthest spins from the locally applied field versus total
polarization in the non-field region, again for field on 3 of 20 and
4 of 20 spins respectively.

The most immediately obvious difference is that if the local magnetic
field is placed on an odd number of spins, neither spin sequestration
nor closeness in trace distance to the singlet subspace for any spins
are observed, except for the L=0 subspace in weak local fields. Figs.
\ref{fig:periodic SDvEM}(a) and (c) show the trace distance from
a singlet in spins far from the applied magnetic field and local angular
momentum for spins far from the local magnetic field respectively,
both versus total angular momentum in the field free region. For larger
fields, the spins far from the region where the external magnetic
field is applied do not approach the singlet subspace because of frustration
caused by having an odd number of spins in the non-field region. In
the case of periodic boundary conditions, the effects of the frustration
are stronger than in the case of open boundaries. This is because
here a change in the length of the field-to-far-from-field transition
region will do nothing to relieve the frustration because of the symmetry
between the two field boundaries. Regardless of whether the length
of one of these regions is odd or even, the total length of transition
regions is always even because it is the length of a single transition
region multiplied by two.

\subsection{Effect of local degeneracy on small quenches}

\begin{figure}
\includegraphics[scale=0.5]{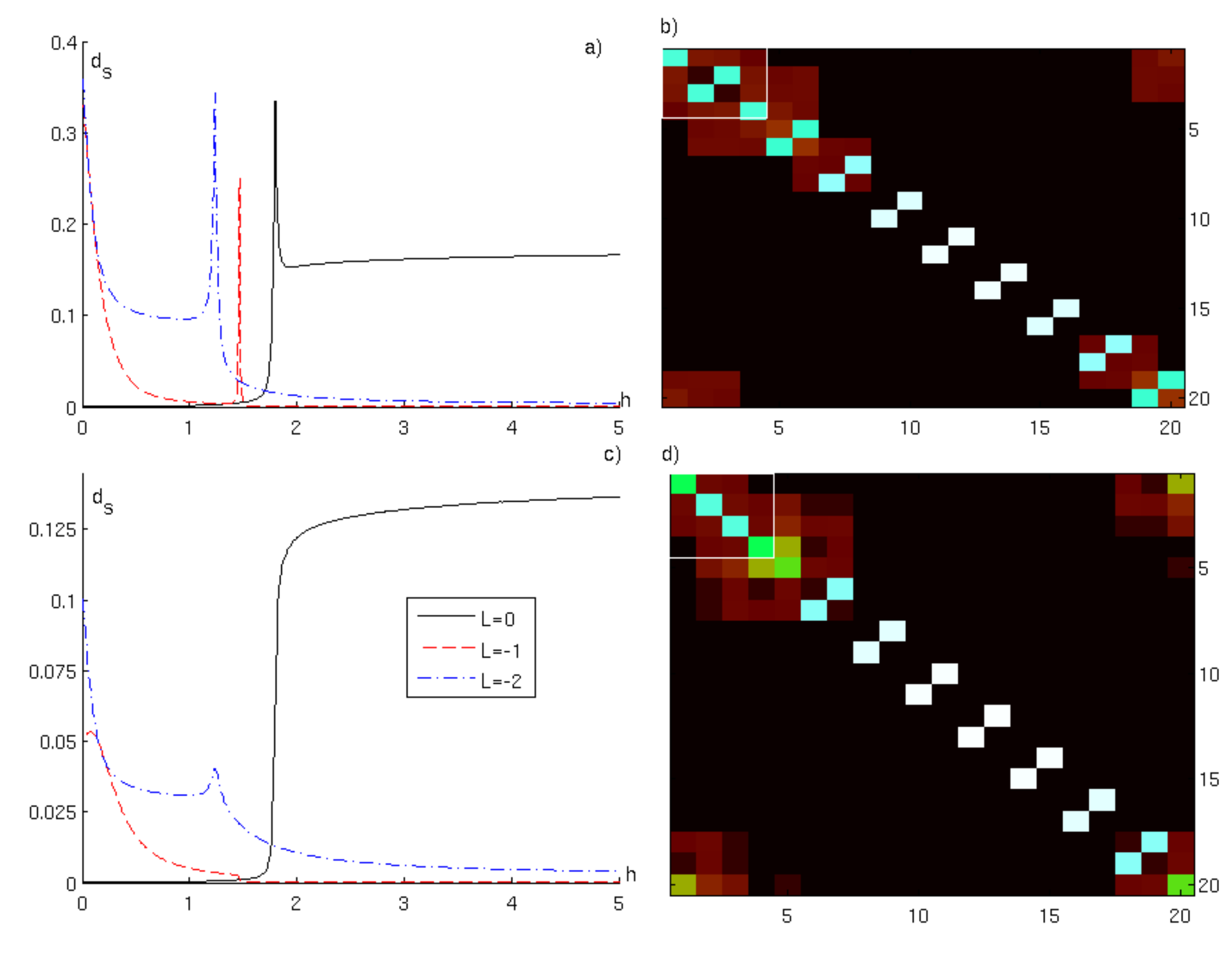}

\caption{(color online) a) Trace distance between the two spins furthest from
the field and the nearest singlet covering (see Eq. \ref{eq:dist sing cover})
versus field for 20 spins with periodic boundary conditions and a
field placed on 4 of the spins. b) Entanglement map for 16 spins with
a field of h=1.3J placed on spins 1-4 (indicated by the white rectangle)
with periodic boundary conditions c) Same as (a), but now with distance
to the closest state in the degenerate subspace (see Eq. \ref{eq:dist sing subspace})
d) Same as (b) but with a field of h=1.6J. On all figures, the solid
line is the L=0 sector, dashed lines indicate the L=-1 sector and
the dot dashed lines indicate the L=-2 sector, where negative L is
in the direction of the field. All plots in this figure are for eignestates
of the Hamitlonian.}

\label{fig:singlet shift} 
\end{figure}

Shifting the focus to the case where the external field is placed
on an even number of spins, one can consider the effects of now having
a locally degenerate ground state, i.e. having a Hamiltonian which
has a ground state degeneracy when no field is applied, and therefore
is degenerate in a local sense far from the spins with an applied
magnetic field. Fist the ground state can be studied by observing
Fig. \ref{fig:singlet shift}, this figure shows in parts (a) and
(c) the trace distance from the closest singlet covering and minimum
distance from the manifold of singlet coverings respectively for the
two furthest spins from the locally applied magnetic field versus
field strength, for a field applied to 4 of 20 spins with periodic
boundary conditions. Parts (b) and (d) show entanglement maps for
a local magnetic field strength of h=1.3J and h=1.6J respectively,
again for a field on 4 of 20 spins with periodic boundaries. Fig.
\ref{fig:singlet shift}(c) indicates that the global ground state
of the system is always close to the singlet subspace far from the
field, however \ref{fig:singlet shift}(a) suggests that around a
field strength of 1.5 the system may undergo a switch between singlet
coverings far from the spins to which the field is applied. Figures
\ref{fig:singlet shift}(b,d) confirm this suspicion by showing that
indeed before the peak in \ref{fig:singlet shift}(a) there are an
even number of dimers outside of the field region, while after there
are an odd number of dimers. This indicates that disturbances from
the local field can be felt far from the spins with an applied field,
but only for a narrow range of field values.

\begin{figure}
\includegraphics[scale=0.5]{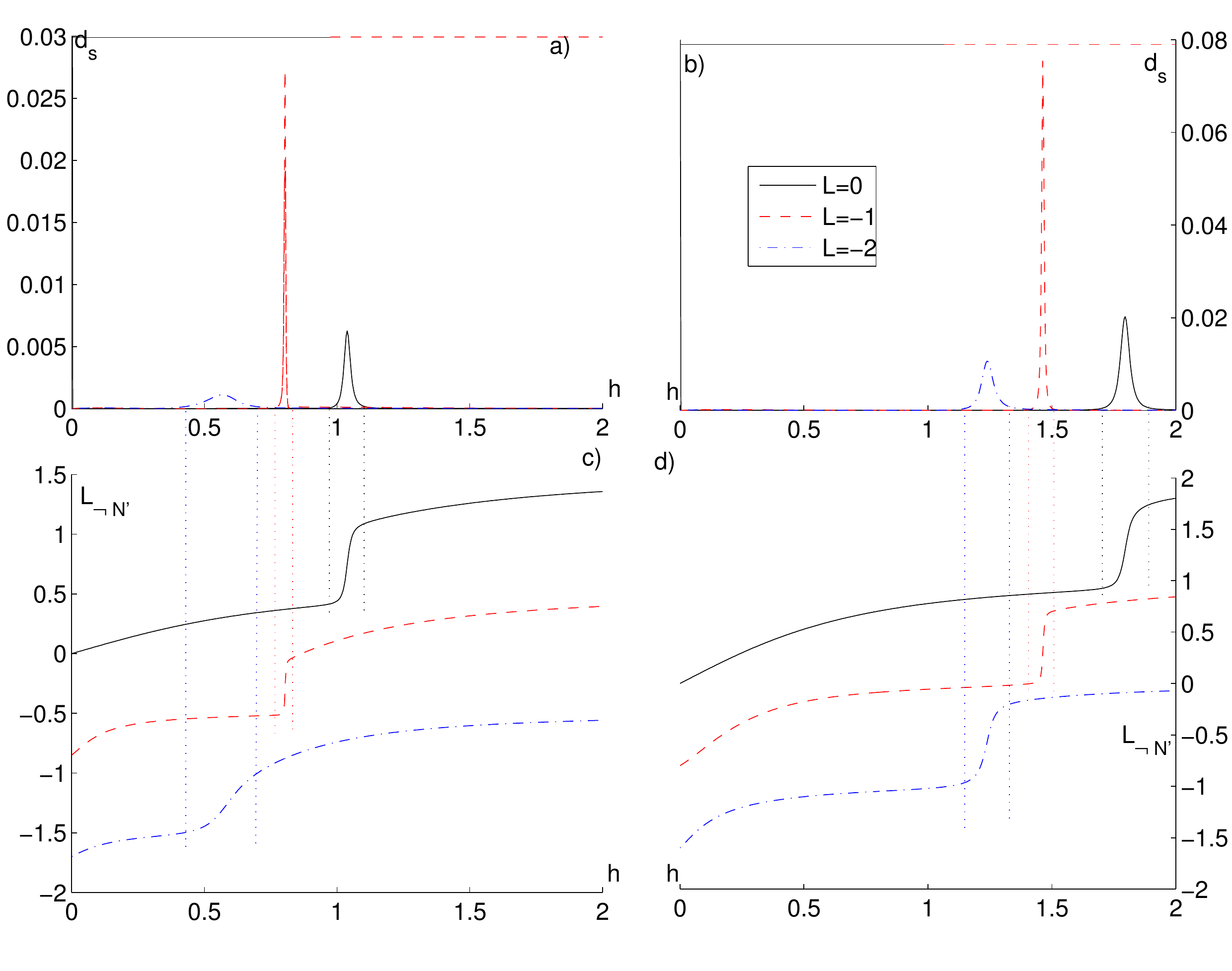}

\caption{(color online) Initial trace distance from average (see Eq. \ref{eq:dist av})
for a subsystem far from the fields after a small field quench $\epsilon=0.001$
(top row). Polarization on all non-field spins versus field strength
(bottom row). 20 spins with field placed on 3 of them and periodic
boundary conditions (left column). same with field placed on 4 spins
(right column). Dotted vertical lines have been added to emphasize
correlation between the two graphs. On all figures, the solid line
is the L=0 sector, dashed lines indicate the L=-1 sector and the dot
dashed lines indicate the L=-2 sector. Lines at the top are added
to show which spin sector the global ground state is in. The top two
plots are time averaged quantities from a quench, while the bottom
two figures are properties of the ground state of each sector.}

\label{fig:initial dist to av} 
\end{figure}

One can now consider the effect of small quenches at various applied
field strengths on spins far from the field spins. The results of
such quenches are shown in Fig. \ref{fig:initial dist to av}, parts
(a) and (b) show the trace distance to average for the two furthest
spins from the locally applied magnetic field, after a quench which
involves a small change in field strength versus the strength of that
field for a field on 3 of 20 and 4 of 20 spins respectively with periodic
boundary conditions. Parts (c) and (d) show the polarization of the
two furthest spins from the locally applied magnetic field versus
field, and are included to emphasize the important role played by
polarization in this system. One would expect that these disturbances
can only be propagated through the still locally degenerate ground
state subspace of the no-field Hamiltonian and therefore would only
have an effect when the coverings shift. Fig. \ref{fig:initial dist to av}(b)
shows that in fact a small quench does disturb the system strongly
at the point where the coverings switch. The other two peaks in Fig.
\ref{fig:initial dist to av}(b) are less relevant because they occur
in the ground state of a spin sector, but not in the global ground
state of the system. Also none of the quench disturbances which occur
far from the spins with an applied field occur in the global ground
state in Fig. \ref{fig:initial dist to av}(a), demonstrating another
difference caused by even-odd effects. This is to be expected, because
the the two degenerate ground states of an odd length Majumdar-Gosh
chain lie in different polarization sectors and therefore cannot exhibit
level repulsion, at least locally, in the region far from the applied
magnetic field.

Note also that Fig. \ref{fig:initial dist to av} suggests that there
is a strong correlation between polarization outside of the subsystem
where a magnetic field is applied and quench disturbance to the far
spins, in the sense that when the quench has a strong effect, there
is a rapid change in polarization in the non-field region. The converse
however is not supported by this figure. This demonstrates than polarization
plays a strong role in the global behavior of this system.

The same energy arguments used in the static case for behavior of
spins far from the spins with an externally-applied magnetic field
should be usable as a dynamical argument. A finite local field can
only introduce a finite amount of energy into the system. Therefore
only states which lie sufficiently close in energy to the ground state
can be accessed in any significant way. For a large enough system,
all of the low energy states will have to be locally close to the
ground state for most of the spins far from the locally-applied magnetic
field, therefore locally, far from the field, spins can only be disturbed
within the degenerate subspace. Put another way, in gapped systems
the effects of a local quench have to be localized, unless there exists
a locally degenerate subspace far from the region where the quench
is applied. When such a subspace exists it may be able to transport
conserved charges, quantum entanglement and dynamical disturbances
an unlimited distance away from the disturbance site. A locally degenerate
ground state can be thought of as a special symmetry which allows
transport of information and charges (but not energy) with no losses
throughout the part of a system far from the quench. \cite{end7}

\begin{figure}
\includegraphics[scale=0.5]{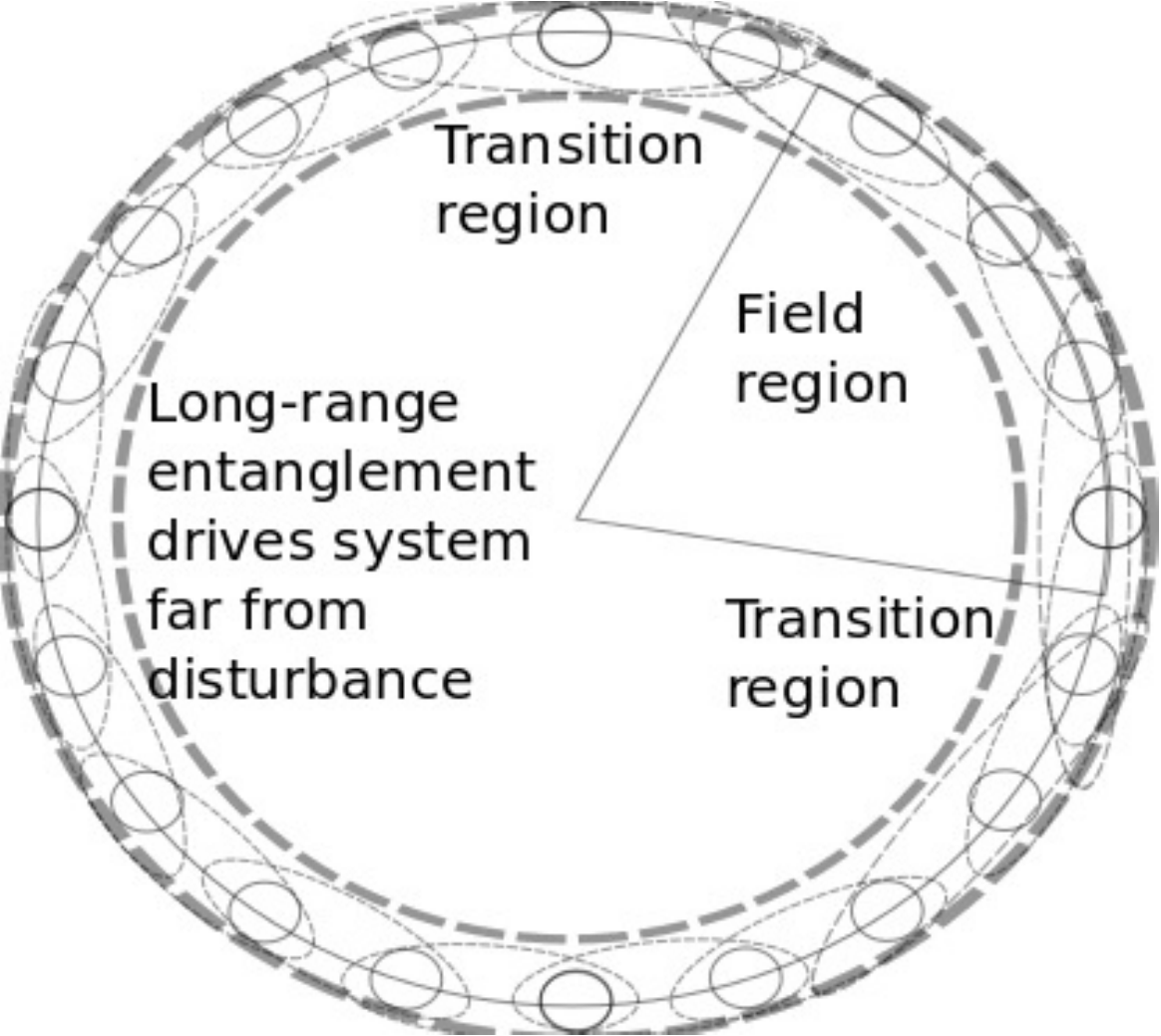}

\caption{Cartoon of field quench for periodic boundaries.}

\label{fig:long range cartoon} 
\end{figure}

Long-range entanglement allows a part of the system which lies entirely
in a degenerate subspace to have its evolution driven by local evolution
far away, see Fig. \ref{fig:long range cartoon}.

\subsection{Large quenches}

\begin{figure}
\includegraphics[scale=0.5]{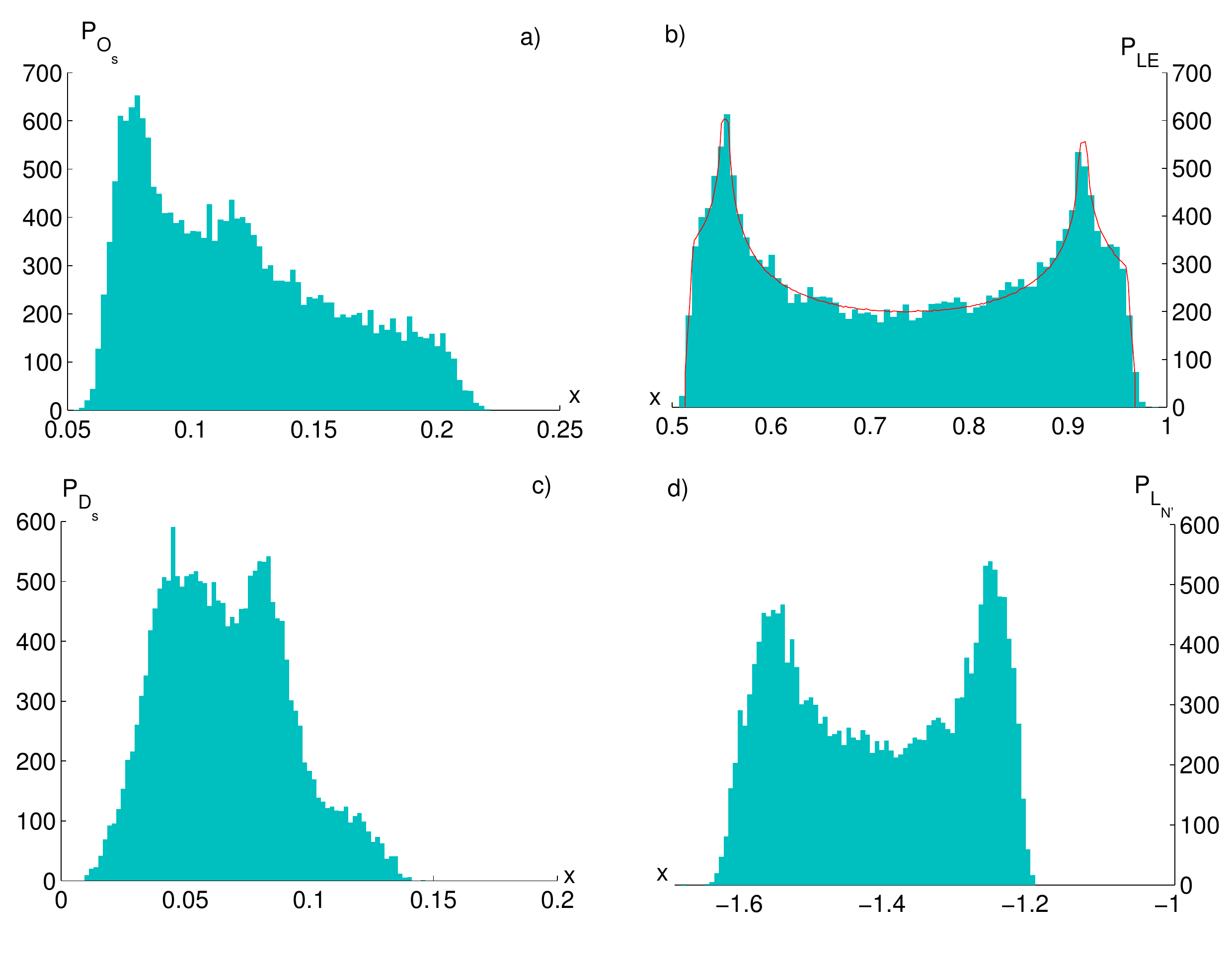}

\caption{(color online) Equilibration statistics for N=20, N'=4 with a quench
from h$_{(i)}$=1.6 to h$_{(f)}=1.3$. a) Time statistics of the trace
distance of 4 spins far from the locally applied magnetic field from
a singlet state (Eq. \ref{eq:singlet overlap}). b) Time statistics
of the Loschmidt echo (Eq. \ref{eq:Loschmidt}) with an approximation
based on few frequencies. c) Time statistics of distance from time
averaged state (Eq. \ref{eq:time dist av}) for 4 spins far from the
locally-applied magnetic field. d) Time statistics of local magnetization
(Eq. \ref{eq:locMag}) of the field spins. These plots are all time
statistics obtained from evolution.}

\label{fig:large quench statistics} 
\end{figure}

Now that it is established that a disturbance will be able to be propagated
throughout the entire system from a small quench, one can perform
a large quench from h=1.6 to h=1.3 for a local field applied to 4
adjacent spins of 20 total spins with periodic boundary conditions.
One can then examine the time statistics of various properties of
the system. These statistics are shown in Fig. \ref{fig:large quench statistics},
in this figure part (a) is the trace distance of 4 spins far from
the locally applied magnetic field from a singlet state, part (b)
is the time statistics of the Loschmidt echo of the entire system,
part (c) is the time statistics of the distance from the time averaged
state for 4 spins far from the locally applied field, and part (d)
is the time statistics of the magnetization of the spins subjected
to the field. These statistics will show the ability of the system
to equilibrate, even locally for spins far from the spins where the
local magnetic field is applied. In the case studied here, the system
only equilibrates poorly, even in the global sense, not surprisingly,
poor equilibration is also shown in local observables both close to
and far from the spins with an applied magnetic field.

The double-peaked pattern of equilibration seen here is typical of
small systems, see \cite{Diez2010}, and is thus consistent with the
theory that although the system itself is rather large \cite{end8},
the actual evolution is only taking place on a few spins in or near
the region of externally applied field, the rest of the system is
simply being drug along by long range entanglement with these spins.
As Fig. \ref{fig:large quench statistics}(c) demonstrates, even though
the dynamics is driven by long range entanglement with far away spins,
a subsystem of spins is still able to be pushed toward equilibration
in the trace distance sense. The fact that there is no local energy
difference does not seem to interfere at all with equilibration of
these spins. The trace distance from the average is observed quite
close to zero at some times, unlike in similar quenches performed
at the Majumdar-Ghosh point in \cite{Diez2010}. This is because an
undisturbed singlet somewhere in the region being observed would yield
a large distance from the average at all times as shown in \cite{Diez2010}
where the quench did not cause a change in singlet coverings. In the
case we are observing, where the coverings switch, there are no undisturbed
singlets in the region away from the field spins.

Although the equilibration is globally poor in this system, there
are no signs that equilibration via long-range entanglement through
a locally degenerate subspace is any less effective than direct equilibration
of the spins to which the field is applied. The data from this quench
therefore indicate that the entire system can be equilibrated (at
least somewhat) by a quench which only affects a very small region.
In fact a system of any size should be able to be brought locally
close to equilibrium in this way. Because all states of the far spins
locally have the same energy, than they cannot affect the time evolution
of the system, therefore the same behavior would be expected for a
spin chain of any sufficiently long (even) length.

\section{Other Coupling Strengths}

One can now ask what would happen if the coupling were changed such
that the system was no longer using the Majumdar-Ghosh Hamiltonian,
but allowed the next nearest neighbor coupling to take on arbitrary
values, see Eq. \ref{eq:J2 hamiltonian}. This study is done with
20 spins and periodic boundary conditions, with a local magnetic field
on 4 adjacent spins.

\begin{equation}
H_{J2}=\sum_{j=1}^{N}\vec{S}_{j}\cdot\vec{S}_{j+1}+J_{2}\sum_{j=1}^{N}\vec{S}_{j}\cdot\vec{S}_{j+2}\label{eq:J2 hamiltonian}
\end{equation}

\begin{figure}
\includegraphics[scale=0.45]{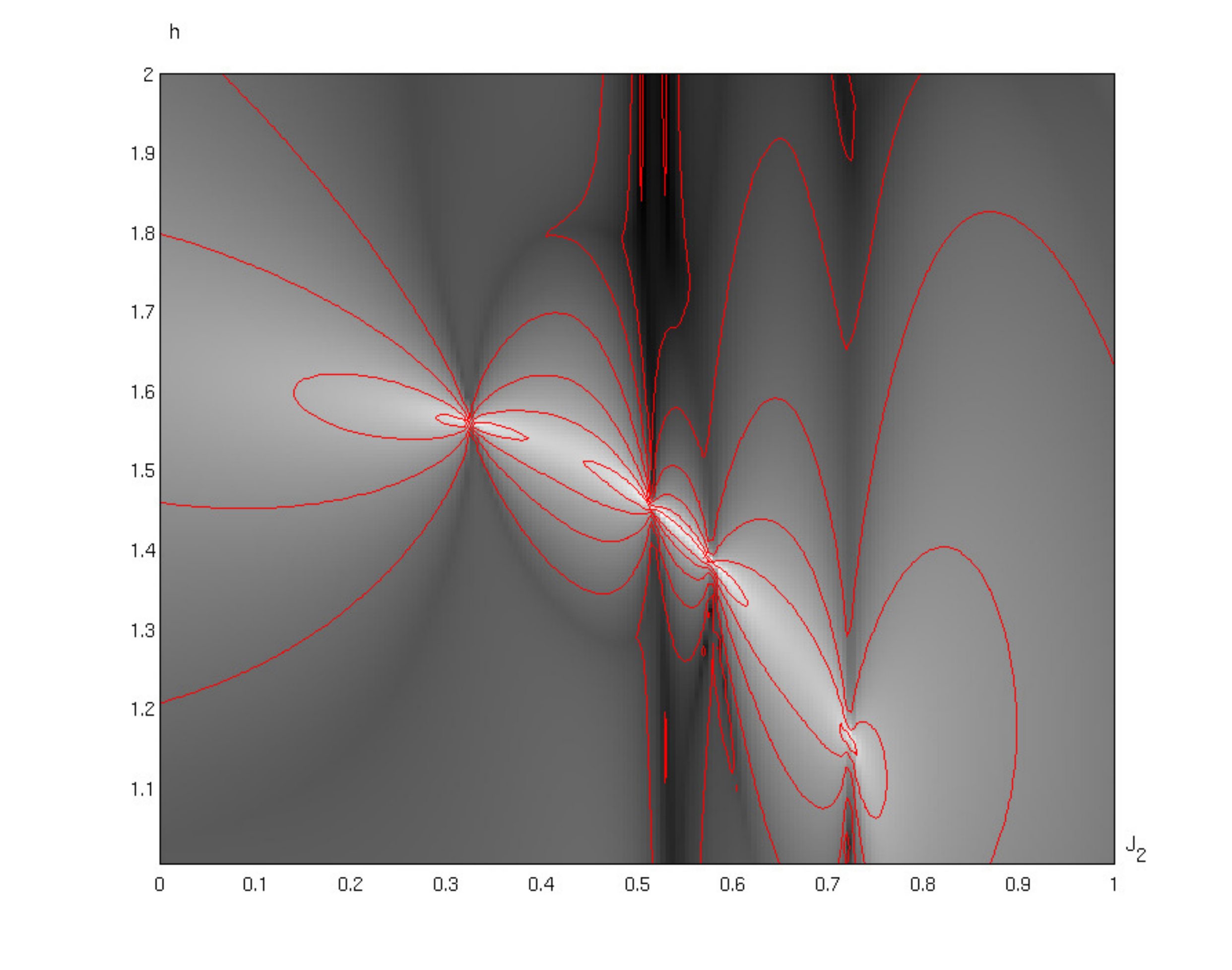}

\caption{(color online) Initial trace distance to average for far spins after
a small field quench, larger distances are lighter, smaller distances
are darker. Trace distance is plotted on a logarithmic scale, contour
lines (red on-line) are included for clarity. Data using 20 spins
with periodic boundaries in the L=-1 sector. \label{fig:Init2avJ2}}
\end{figure}

Small field quenches can be considered on this new Hamiltonian exactly
in the same way they can be considered for the Majumdar-Ghosh Hamiltonian,
the results appear in Fig. \ref{fig:Init2avJ2} which shows the initial
trace distance from average for a small field quench versus coupling
and field strength. This data shows that for a wide range of coupling
near the Majumdar-Ghosh point, small field quenches can drastically
affect spins very far from the spins with an applied magnetic field
at specific field strengths. However, when the field strength is eventually
different enough, these peaks broaden out and disappear (note logarithmic
scale in Fig. \ref{fig:Init2avJ2}). The basic behavior seen previously
in this paper holds for a wide range of couplings, where the ground
state is no longer degenerate.

For an infinite system one would expect that, far spins from the local
magnetic field could not be disturbed by a local field quench unless
either the system is gapless or there exists a degenerate ground state.
For a finite system, this would only be necessarily true if the gap
between the ground state and first excited state is sufficiently large
compared to the energy introduced by the applied local magnetic field,
in which case the field will be unable to introduce enough energy
to affect the entire system. In this case the order of magnitude of
the energy which the field introduces can be estimated by simply multiplying
the field strength by the number of spins it is applied to. Because
both of the quantities are of order 1, one would expect that the energy
introduced would also be of order 1.

\begin{figure}
\includegraphics[scale=0.45]{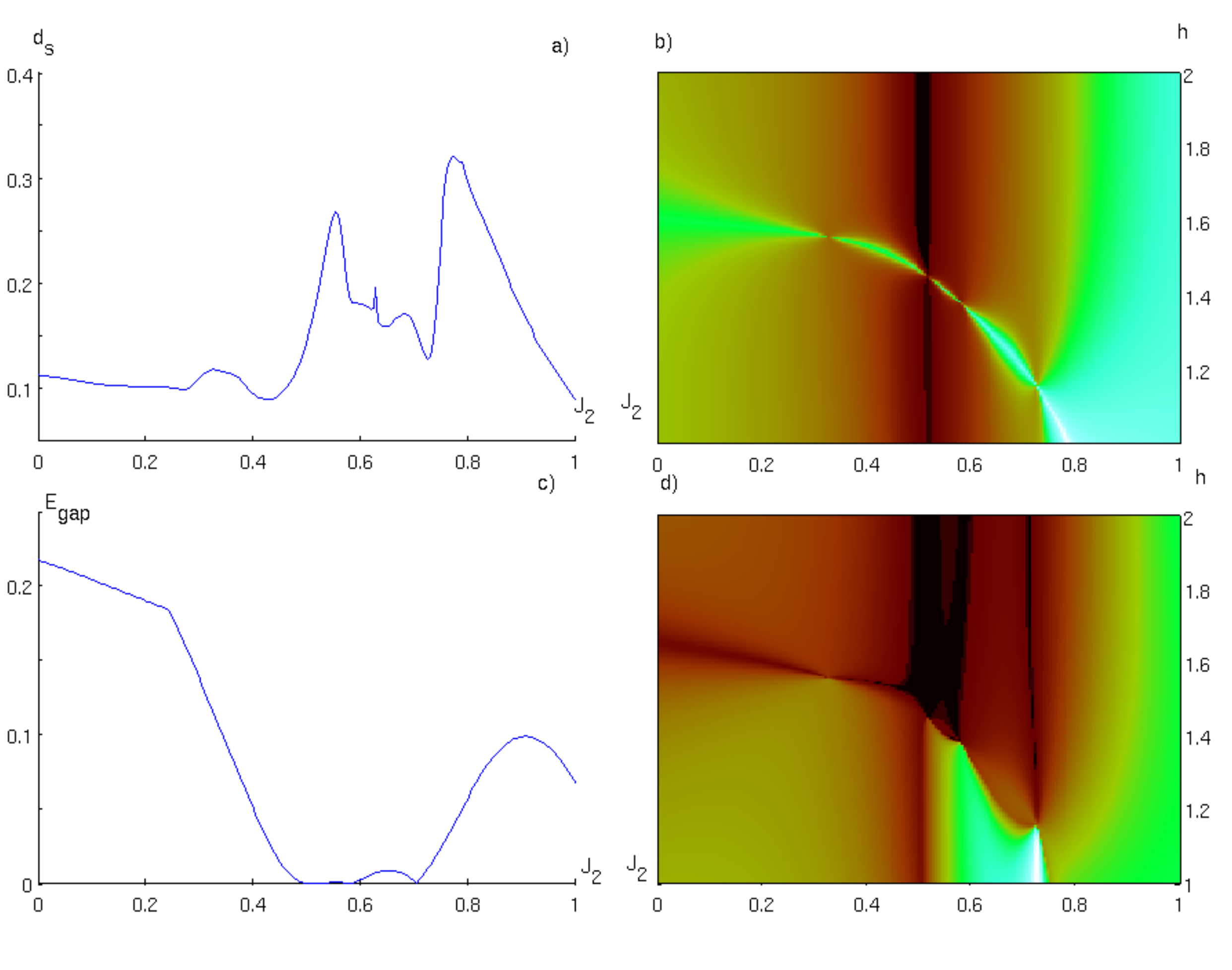}

\caption{(color online) a) Initial distance from average for 4 spins far from
the locally applied field for a large quench from h=2 to h=1 for the
L=-1 sector. b) distance of far spins from nearest singlet covering
(Eq. \ref{eq:dist sing cover}) versus h and $J_{2}$, L=-1 sector,
color scale same as for entanglement maps, but normalized to largest
value. c) gap between ground state energy and first excited state,
for different $J_{2}$and h=0. d) distance from singlet subspace for
far spins (Eq. \ref{eq:dist sing subspace}) versus h and $J_{2}$,
L=-1 sector, color scale same as for entanglement maps, but normalized
to largest value. All plot except for (a) are static quantities relating
to eigenstates.}

\label{fig:differentJ2} 
\end{figure}

The energy gap in the system which will be used for this calculation
can be determined by exact diagonalization. The energy of the gap
between the ground state and first excited state of this system are
shown in Fig. \ref{fig:differentJ2} part (c) which shows the gap
energy versus coupling at zero applied field, part (a) shows the initial
distance from the average for 4 spins far from the locally applied
magnetic field after a large quench (within the $L=-\frac{1}{2}$
sector), part (b) shows the local trace distance from the nearest
singlet covering for spins far from the local field in the ground
state of the $L=-\frac{1}{2}$ sector versus field strength and coupling,
and part (d) is the same as (b), but with trace distance from the
nearest state in the ground state manifold. It can be seen from Fig.
\ref{fig:differentJ2}(c) that the gap energy is at most of order
0.1, therefore, one would expect that for the entire range of couplings,
the far spins could be disturbed by the local field. The results seen
in Fig. \ref{fig:Init2avJ2} are as expected, however if the system
size were increased to infinity, one would expect that in the gapped
region for $J_{2}\gtrsim$0.25, the peaks in the distance would have
to disappear except for exactly at $J_{2}$=0.5, or any other point
with a degenerate ground state. Twenty spins, however, is still too
small a system for changes in the coupling to destroy the ability
to dephase far spins with a local field, in other words the system
can be considered to have an approximately degenerate (wrt. the energy
scale associated with the field) ground state for all values of $J_{2}$,
the next nearest neighbor coupling.

One can now ask whether the effects seen in Fig. \ref{fig:Init2avJ2}
away from the Majumdar-Ghosh point are also caused by some kind of
shift in singlet covering. To answer this question, one can compare
Fig. \ref{fig:differentJ2}(b) to Fig. \ref{fig:differentJ2}(d) and
notice that where the peaks in Fig. \ref{fig:Init2avJ2} are located,
the trace distance from either covering tends to be relatively large,
but the distance from the subspace tends to be relatively small. This
indicates that movement within the singlet subspace is the cause of
much of the disturbance in the far spins. Also interesting to note
is that for a significant portion of the couplings, the spins far
from the locally applied magnetic field are closest to the singlet
subspace when the small field quenches have the most effect on far
spins. It appears that even at many couplings away from the Majumdar-Ghosh
point, the model of switching between coverings as a way to spread
a disturbance throughout the system is accurate. In fact for many
values of $J_{2}$, the system appears to move into the singlet subspace
for a narrow range of fields only when the covering change occurs.
For $J_{2}\gtrsim0.6$ this model seems to break down, but it is still
at least relevant for a large range of $J_{2}$. Although not directly
related to the quench, it is interesting to note that above a certain
local magnetic field strength the spins far from the field seem to
lie on the singlet superposition manifold for a fairly large range
of coupling strengths near Majumdar-Ghosh coupling, as well as a narrow
strip between $J_{2}=0.6$ and $J_{2}=0.8$, the reason for this is
not known.

The results of a large local magnetic field quench over varying $J_{2}$
as shown in Fig.\ref{fig:differentJ2}(a) simply helps to underscore
what has already been noted about changing coupling not being an effective
way of preventing disturbances from propagating throughout the system
at this system size. Not only do the large quenches have a significant
effect on far spins from the local magnetic field for all coupling
strengths, but the expected trend of decreasing quench effect with
increasing gap is not visible in any definitive way, indicating that,
not only is the energy scale of the gap (Fig. \ref{fig:differentJ2}(c))
too small to be the dominating factor in the quench effectiveness,
it seems to not even play a very significant role. This result is
consistent with the previous energy scale argument, the energy scale
associated with the field is always at least an order of magnitude
larger than the gap between the first 2 eigenstates.

\section{Conclusions}

In systems with degenerate ground states, quantum entanglement, disturbances,
and charges can propagate freely, as long as the quench crosses between
pre and post quench ground states which are locally different from
each other far away from the region affected by the local quench.
This effect is different and independent from gapless excitations,
and has been demonstrated to occur in a gapped system. Unlike in gapless
systems where excitations carry an arbitrarily small amount of energy
far from the quench, these excitations store all energy locally near
the quench, and evolution far away is driven solely by long-range
entanglement. The local energy far from the region affected by a local
quench Hamiltonian is exactly zero in these systems, not arbitrarily
small.

To allow a charge to be propagated through a degenerate subspace,
the two degenerate ground states must have different local expectation
values for said charge far from the region affected by a local quench
Hamiltonian. An effectively odd spin chain far from the field is allowed
to propagate polarization throughout the far region for example. Again,
in such a case, long range entanglement can propagate the charge,
but does not propagate any energy far from the field. In cases where
two degenerate ground states with different expectation values for
a charge far from the region where the quench is applied do not exist,
the charge can become locally trapped in part of the system. In the
case of the Majumdar-Ghosh Hamiltonian, the polarization is trapped
near the boundary of the local magnetic field region. A locally unique
ground state (in a gapped system) means that charges, as well as disturbances,
are confined after a local quench. Energy arguments prevent a disturbance
from traveling throughout the system and also therefore forbid charges
from moving outside of a small area.

In the system studied here, a large local magnetic field causes the
spins within the field to become effectively 'fixed', facing in the
direction of the field in the ground state. An approximation which
does not include these spins directly but includes an effective modulation
in coupling between the two spins neighboring the field can faithfully
reproduce the ground state when an odd number of spins remain. For
the case that an even number of spins are left out of the field region,
this simple approximation fails. We believe that the effective transition
region between the field and non-field region consists of an odd number
of spins, and that this ground state cannot be faithfully reproduced
in this way because of odd length frustration effects.

For Majumdar-Ghosh spin chains with periodic boundaries, with a local
magnetic field on some even number of spins, there exists a range
of fields where a small field quench can propagate a disturbance through
the entire system using long range entanglement. This disturbance
is propagated locally through the degenerate subspace of the local
ground state. In this case, this range of fields is relatively narrow.
A quench across this entire range does not only cause equilibration
near the field, but also moves the far spins towards equilibration,
within the locally degenerate subspace.

Study of systems with different next nearest neighbor coupling indicate
that the basic effect which causes disturbances to be propagated to
far spins can, at least for small enough systems, be extended away
from the Majumdar-Ghosh point. For finite systems, if the gap between
the ground state and the first excited state is small enough, the
same effect which was described here for degenerate systems can also
be applied to systems where the first 2 states are close in energy.
In other words, under the right conditions, an approximate degeneracy
will work in place of an exact degeneracy. For a system of 20 spins
any value of $J_{2}$ between 0 and 1 still allows the far spins to
be significantly affected by the field. We strongly suspect that for
the values of $J_{2}$where the Hamiltonian is gapped and for which
a degenerate ground state does not exist, spins far from a local magnetic
field applied to a few spins cannot be affected in the large system
limit.

\subsection*{Acknowledgements}

The authors would like to thank N. Tobias Jacobson for his assistance
in editing this paper. We would also like to thank S. Garnerone, B.
Normand, and M. Diez for valuable discussions. The numerical computations
were carried out on the University of Southern California high performance
supercomputer cluster. This work a has been supported by NSF grants:
PHY-803304,DMR- 0804914.


\begin{thebibliography}{References}
\bibitem{end1}However, one can orthogonalize them, still preserving
the symmetry under $a\longleftrightarrow b,\,2l\rightarrow mod_{N}(2l+1)$,
by adding a term $-(a\times\bigotimes_{k=1}^{\frac{N}{2}}\frac{\ket{\uparrow_{2k-1}\downarrow_{2k}}}{\sqrt{2}}+b\times\bigotimes_{k=1}^{\frac{N}{2}}\frac{\ket{\downarrow_{2k-1}\uparrow_{2k}}}{\sqrt{2}})$.
\bibitem{end2}The only case where this is not observed is for the
case of periodic boundary conditions with a local field applied to
an even number of spins. In this case these spins orient locally like
in the frustrated ground state of a Majumdar-Ghosh chain with an odd
number of spins.
\bibitem{end3}Note that due the inhomogeneity introduced by the applied
local field, translational symmetry is broken.
\bibitem{end4}An example of this would be to consider a superposition
of the two singlet coverings. Any non-adjacent spins will have exactly
zero two-point entanglement (see Eq. \ref{eq:entMap}). However, any
set of two pairs of adjacent spins will have a finite entanglement
between them. To see this, consider the case where one set of two
spins is measured to be both in the up direction. If there is an odd
number of spins between the two spin pairs, this forces the other
pair to be a singlet, regardless of the distance between the pairs.
\bibitem{end5}Showing that this happens consists of demonstrating
that the singlet covering is an eigenstate of any field which does
not change within a singlet. This can be seen by realizing that the
field operator on the first spin of the singlet will give the zero
magnetization triplet state $\frac{1}{\sqrt{2}}(\ket{\uparrow\downarrow}+\ket{\downarrow\uparrow})$,
while the operator on the second spin will give the same but with
a negative sign. Thus the two will cancel making the singlet covering
an eigenstate with a zero eigenvalue.
\bibitem{end6}The closeness to the boundary may also be a factor
in the trace distance of the last 2 spins from a singlet, the fact
that no entanglement can cross the open boundary may cause spins close
to it to assume more localized states, however while this effect could
make finite distances smaller, it should not be able to make the trace
distance (virtually) zero as it is in many parts of Fig. \ref{fig:sd_&_L_vs_h}(a).
\bibitem{end7}Note that the excitations which travel through a locally
degenerate ground state are not the same as gapless excitations, which
locally carry an arbitrarily small amount of energy. The parts of
these excitations which exist far from the region affected by the
local quench Hamiltonian carry exactly zero energy locally.
\bibitem{end8}One could argue that 20 spins is not such a large system,
but it has been shown that the spins far from the evolution are always
locally in the degenerate subspace. Adding more far spins and making
the system into one which all of the readers would agree would be
{}``large'' (for example making the system size 100,000 spins) would
not effect the dynamics, and the double peaked pattern would remain.
\end{thebibliography}
\end{document}